\documentclass[conference]{IEEEtran}
\IEEEoverridecommandlockouts
\usepackage{cite}
\usepackage{amsmath,amssymb,amsfonts}
\usepackage{algorithmic}
\usepackage{graphicx}
\usepackage{textcomp}
\usepackage{xcolor}
\usepackage[colorlinks=true, allcolors=blue]{hyperref}
\usepackage{float}
\usepackage{tabularx}
\usepackage[english]{babel}
\usepackage{enumitem}
\usepackage{tabularx}

\newlist{steps}{enumerate}{1}
\setlist[steps, 1]{label = Step \arabic*:}
\def\BibTeX{{\rm B\kern-.05em{\sc i\kern-.025em b}\kern-.08em
    T\kern-.1667em\lower.7ex\hbox{E}\kern-.125emX}}
\begin{document}


\title{Causal Intervention Sequence Analysis for Fault Tracking in Radio Access Networks}

\author{
\IEEEauthorblockN{Chenhua Shi, Joji Philip, Subhadip Bandyopadhyay, Jayanta Choudhury}
\IEEEauthorblockA{Ericsson \\
\texttt{\{chenhua.shi, joji.a.philip, subhadip.bandyopadhyay, jayanta.choudhury\}@ericsson.com}}
}

\IEEEpubid{%
\makebox[\columnwidth]{979-8-3315-5397-5/25/\$31.00~\copyright~2025 IEEE \hfill}%
\hspace{\columnsep}%
\makebox[\columnwidth]{ }%
}

\maketitle
\IEEEpubidadjcol

\begin{abstract}
To keep modern Radio Access Networks (RAN) running smoothly, operators need to spot the real-world triggers behind Service-Level Agreement (SLA) breaches well before customers feel them. We introduce an AI/ML pipeline that does two things most tools miss: (1) finds the likely root-cause indicators and (2) reveals the exact order in which those events unfold. We start by labeling network data: records linked to past SLA breaches are marked `abnormal', and everything else `normal'. Our model then learns the causal chain that turns normal behavior into a fault. In Monte Carlo tests the approach pinpoints the correct trigger sequence with high precision and scales to millions of data points without loss of speed. These results show that high-resolution, causally ordered insights can move fault management from reactive troubleshooting to proactive prevention.
\end{abstract}


\section{Introduction}
As network complexity increases, diagnosing and mitigating faults within radio access networks (RANs) have become more challenging. Traditional supervised machine learning models often rely on low-resolution, aggregated data (e.g., 15-minute granularity) and necessitate human intervention for ground truth annotation. In addition, data preparation with a proper ground truth requires a thorough analysis and review by human experts. A critical hindrance to this approach in telecom is the involvement of human experts in labeling the data for ground truth. The analysis of milliseconds and sub-milliseconds granularity of real-time (RT) data is slow and cumbersome for human experts, as such data are collected in bulk with massive amount of trace and error logs. Since 15-minute granularity performance data are aggregated from minutes, seconds, and milliseconds granularity data, the volume of high-resolution data explodes. It is imperative to execute computationally intensive AI/ML algorithms on the seconds and milliseconds granularity high-resolution data to isolate the root cause candidates and apply remedy prior to the occurrence of SLA violation on the 15-minute granularity data. This paper proposes an automated high-resolution analysis framework, capable of identifying the order of root cause of SLA violations by using Root Cause Discovery algorithm(s), causal subgraph construction, and deviation detection algorithms.

The paper is organized in the following sections. Section II provides existing AI/ML approaches for troubleshooting RAN issues. Our approach, combining three functional components to determine the causal intervention sequences or deviation events leading up to the SLA breaches is elaborated in Section
III. In Section IV, we introduce the results of the experiment to identify the potential root cause of cell load issues using the PCMCI technique and our approach. We then showcase how to determine the best choice of parameters for the Root Cause Discovery algorithm under Monte Carlo simulation. Finally, we conclude with closing remarks in the subsequent section.

\section{Related Work}
Early AI/ML tools for RAN fault diagnosis \cite{alan2024} \cite{antor2024} \cite{Qiuping2021} \cite{HNN} rely on 15-minute KPI aggregates. At that resolution, noise is smoothed out and manual feature selection is still manageable. By contrast, sub-second data are both enormous and volatile. No engineers can sift through the flood of metrics pouring in from thousands of cells, and building a clean, labeled ground-truth set for supervised learning is—at scale—unrealistic.

The research community is actively exploring ways to provide aid by automation using AI/ML methods.
Among the relevant works, the development of the Differentiable Adjacency Test (DAT-Graph method) \cite{alan2024} is suitable for scalable causal discovery with both observational data and soft intervention data, indicating changes in the distribution profile of random variables. However, it is not well suited for discovering root causes.

To automate insight extraction, researchers have turned to causal-discovery algorithms. The Differentiable Adjacency Test (DAT-Graph) \cite{alan2024} is a landmark example: it scales to large graphs and handles soft interventions (i.e., distribution shifts). Unfortunately, it assumes that each variable keeps a stable distribution over time. Hard interventions—when a KPI gets pinned to a constant during a fault—break this assumption, so DAT-Graph often misses the real culprit. Variants such as PCMCI\cite{runge2019}, PCMCIplus\cite{runge2020}, LPCMCI\cite{gerhardus2020}, RPCMCI\cite{saggioro2020} inherit the same limitation; structural-causal theory calls these do-interventions \cite{pearl2009}. In practice, an operator needs two things:

\begin{enumerate}
\item A shortlist of root-cause indicators that truly matter.
\item The temporal chain that links those indicators to the eventual SLA breach.
\end{enumerate}

Existing algorithms tend to excel at one piece but not both. Methods \cite{runge2019} \cite{peter2013} \cite{azam2022} that build a causal graph under normal, stationary conditions struggle once faults disrupt those links. Conversely, pattern-mining tools that flag anomalies rarely explain why events unfold in a particular order. 

\section{Methodology}

\subsection{Summary}
In order to determine the causal intervention sequences or deviation events leading up to the SLA breaches, we propose an approach combining three functional components. Each component uses AI/ML algorithms to attain a specific outcome that contributes to the primary goal. During the occurrence of fault, variables start to deviate from normal behavior and become either an effect or a cause of intervention. In our solution, a dataset preceding the SLA violation event and a dataset from post SLA violation event are provided as input to the solution pipeline. The first component identifies the variables that have undergone intervention and are causally related to the main indicators of SLA during normal operation. RCD algorithm(s), described in \cite{azam2022} is the essential module in this component. All subsequent references to RCD pertain to this approach described in \cite{azam2022}. It applies causal discovery to identify root cause candidates and uses a modified PC (Peter-Clark)  algorithm under the hood. The algorithm then identifies the intervention variables among those causally related variables during abnormal states, helping to pinpoint the likely root causes with a significant reduction in delay. The second component determines a new causal subgraph for the intervention variables, such that the causal substructure emphasizes the causal connections between the reduced subset of indicators, consisting of the intervention variables and the variables contributing to the SLA condition, as in \cite{runge2019}. The third component applies univariate deviation detection on the time series data of the selected intervention variables, identified as root cause candidates in the second component to capture the temporal anomaly events comprising the sequence of the intervention incidents that led to the SLA violation and identify the earliest event as the most likely root cause of the given data as in \cite{darling1957} and \cite{edward1968}.

\subsection{Approach}
RCD is applied to capture the intervention indicators as root cause candidates. The causal substructure involving the above found indicators is then determined for stable behavior. Then the pattern of event sequence, tracing the failure in the causal substructure of stable behavior, is determined using different methods such as Kolmogorov-Smirnov (K-S) test \cite{darling1957} and Z-Score \cite{edward1968}. Here is the detailed breakdown of the process:
\begin{enumerate}
  \item Define the SLA rules for abnormal behavior and collect high-resolution data during the abnormal state of the RAN.
  \item Collect high-resolution data for normal state behavior, preceding the abnormal state.
  \item Determine the intervention indicators from the collected data.
  \item Create a causal subgraph, representing the causal substructure of the normal state, involving the indicators found above.
  \item Find the temporal sequence of intervention events in the causal subgraph, defined above, by using Kolmogorov-Smirnov test of statistical inequality and Z-Score.
  \item Determine the likely root cause indicators, according to the reliability of the observed intervention events and the lead time.
\end{enumerate}

RCD employs Initial Partitioning with Localized Learning and Hierarchical Refinement steps to enhance the efficiency of the root cause identification process.

Initial Partitioning and Localized Learning
\begin{enumerate}
\item F-NODE Definition (Binary indicator for hard intervention failure):
\[
F_t = \begin{cases} 
1, & \text{if failure is observed at time } t \\ 
0, & \text{otherwise} 
\end{cases}
\]

\item Causal Conditional Independence Test (Using $\Psi\text{-PC}$ algorithm to identify direct effects of F-NODE)
\[
X_i \perp\!\!\!\perp F_t \mid S, \quad \forall X_i \notin \text{Pa}(F_t)
\]
where \( S \) is a separating set.

\item Score-Based Causal Structure Learning (Optimization of a score function S (G, D) on graph
G and data D):
\[
G^* = \arg\max_G S(G, D)
\]
\end{enumerate}

Hierarchical Refinement
\begin{enumerate}
\item Causal Graph Refinement (Finding parent nodes of F-NODE to refine root cause candidates):
\[
\text{Pa}(F_t) = \{ X_i \mid X_i \to F_t \}
\]

\item Conditional Independence Test for Refinement (Removing false positives):
\[
X_j \perp\!\!\!\perp F_t \mid S, \quad \forall X_j \notin \text{Pa}(F_t)
\]

\item Intervention-Based Invariance (Distributional consistency check across observational and interventional data):
\[
P(X \mid do(F_t = 1)) = P(X \mid F_t = 1)
\]

\end{enumerate}

Kolmogorov-Smirnov Test measures the maximum difference between two cumulative distribution functions (CDFs). We can compare the KPI difference in the normal state and abnormal state based on equation \eqref{KS}.
\begin{equation}
D_n = \sup_x \left| F_1(x) - F_2(x) \right|\label{KS}
\end{equation}
where:
\begin{itemize}
    \item \(D_n\) is the KS test statistic.
    \item \(F_1(x)\) and \(F_2(x)\) are the empirical cumulative distribution functions (ECDFs) of the two samples.
    \item \(\sup_x\) denotes the supremum (maximum) over all values of \(x\).
\end{itemize}

Monte Carlo simulation is a computational technique used to understand the impact of uncertainty and variability in mathematical models and systems. It relies on repeated random sampling to obtain numerical results. We design a Monte Carlo simulation to obtain experimental results that support the theoretical analysis underlying the RCD approach.

RCD includes an input parameter $g$, which determines the number of nodes randomly selected for the causal sub-graph construction. To account for this randomness, the experiment is repeated $n$ times, allowing us to record the frequency with which each KPI is identified as a causal source. Here, for a specific KPI, we define $P_{g,n}$ as the value under the KPI, corresponding to the row where $g$ is the number of nodes and $n$ is the number of iterations of the experiments. $P_{g,n}$ is a binomial random variable with parameters $n$ and $p$, where $p$ is the probability that the KPI is a cause and we want to estimate $p$ using the RCD experiment results. It can showcase the proportion of times the KPI has appeared as a cause in $n$ run of RCD with specific $g$. During the experiments, we want to address below three questions.

\begin{enumerate}
    \item  If the estimate $P_{g,n}$ converges to some value $p$, for all $g$, as $n$ increases.
    \item For a given $g$, how can we compute $p$ from the $P_{g,n}$ values collected from the Monte Carlo simulations.
    \item If we can propose some methods to choose $g$ and $n$ to compute $p$ that is mentioned in the second question.
\end{enumerate}

For a given KPI and a given $g$, following the Lindberg-Feller condition \cite{vijay2015}, if the variance of the estimate of $p$ (the probability of the KPI being a root cause) decreases as $n$ (the number of experiments) increases, then the estimate of $p$ converges to the probability that the corresponding KPI is a causal intervention. Here we use sample proportion of the KPI identified as root cause as the estimate of $p$. Here, we use equation \eqref{ev} to compute the estimated variance where $p$ is the probability that the KPI is a cause and $n$ is the population size that the number of experiments.

\begin{equation}
\sigma = \sqrt{\frac{p (1 - p)}{n}}\label{ev}
\end{equation}

Thus for a given KPI, the experiment is repeated for different values of $g$ and $n$, and we first find the optimal $g$ and $n$ for the KPI in hand as follows:
\begin{enumerate}
    \item Compute median of $p$ and consider corresponding $g$ as the estimated optimal $g$.
    \item Compute proportion of variance reduction, for the chosen $g$, across $n$ = 10, 15, ... , 30, 40, 50. The optimal $n$ is chosen corresponding to max variance reduction from the variance-n curve.
\end{enumerate}

Once we get the optimal $g$ and $n$ for each KPI based Monte Carlo experiment, we want to consolidate and arrive at a single $g$ and $n$ which will work for all the KPIs so that we can run one experiment and arrive quantities of interest, e.g., the probability of being a root cause for individual KPI and/or experimental assessment of consistent behavior of RCD algorithm. The following two formulas, \eqref{ok} and \eqref{on}, do that for us.
  
\begin{equation}
g^* = \max_g \left\{ 
\begin{array}{c}
g \text{ values corresponding to} \\
\text{prominent causal sources}
\end{array}
\right\}
\label{ok}
\end{equation}

\begin{equation}
n^* = \max_n \left\{ 
\begin{array}{c}
n \text{ values corresponding to} \\
\text{prominent causal sources}
\end{array}
\right\}
\label{on}
\end{equation}

\subsection{Benefits}
Our root-cause discovery (RCD) framework watches for subtle timing shifts in key KPIs. When those shifts stray from the normal causal pattern, the system captures a `snapshot' of the offending indicators and arranges them in a hierarchy that mirrors how an engineer would trace the fault. Because the chain of events is explicit, domain experts can review and confirm each step with ease. In addition, it provides several advantages in software configuration, resource consumption, runtime performance, and results validation.

\begin{itemize}
\item Flexible labeling. Engineers can define what counts as an `abnormal' low-resolution period in simple terms (e.g., `availability \textless 99. 5\% for 3 minutes') and let the model drill into the high-resolution data automatically.
\item Any feature set, any size. Unlike deep-learning models that need a fixed input dimension, RCD accepts any mix of KPIs and scales with new metrics as networks evolve.
\item Fast and scalable. RCD samples only a small subset of variables—similar to a random causal forest—so it runs quickly even on data from thousands of cells.
\item Minimal tuning needed. Only 3 hyper parameters (2 for RCD algorithm and 1 for K-S test) are needed for model tuning.
\item Easy to explain and validate. Comparing before-and-after histograms of each causal indicator gives engineers a visual sanity check and supports rigorous statistical testing.
\item RCD needs neither GPUs nor large memory footprints, making it cheap to deploy in micro-services or directly on edge compute devices and is energy efficient for monitoring issues on streaming data.
\item The proposed approach drastically reduces the number of variables to observe for the monitoring of early warning indicators leading to SLA violations.
\end{itemize}

\section{Experiments}
We set up for experiments with PCMCI and RCD to contrast and compare the results to demonstrate the superior performance of the proposed approach. PCMCI can be applied separately on the data corresponding to normal operation and on the data corresponding to abnormal operation. The disparity of the causal structure between the two results can be used to identify the deviation. We show by experiment that if using only PCMCI, the results are much worse than the proposed approach using RCD and anomaly detection using Z-score and K-S test.

\subsection{Dataset Curation}
Real-life network use case data is used for the experiment to demonstrate the results of application of the proposed approach as evidence to support our claims. We focus on the Cell Load Issue in the LTE network, as loading is a very common issue impacting network throughput. We define Downlink Throughput dropping below 500 kbps, as our target SLA breach condition. We have collected different UE and Cell internal events gathered from high resolution performance management events (PM events). And then we calculate KPIs and aggregate them into 15-sec granularity level. Based on the periods of SLA breach occurrences, we extract both normal and abnormal state data from the historical dataset, with each state lasting approximately 30 minutes. For each SLA breach event, this yield 240 data points across 60 KPIs, which are then fed into the root cause discovery algorithms to identify potential intervention variables, as root cause candidates from the normal state and abnormal state data.

\subsection{PCMCI on Cell Load Issue}
\begin{figure}[htbp!]
\centering
\includegraphics[width=0.8\linewidth]{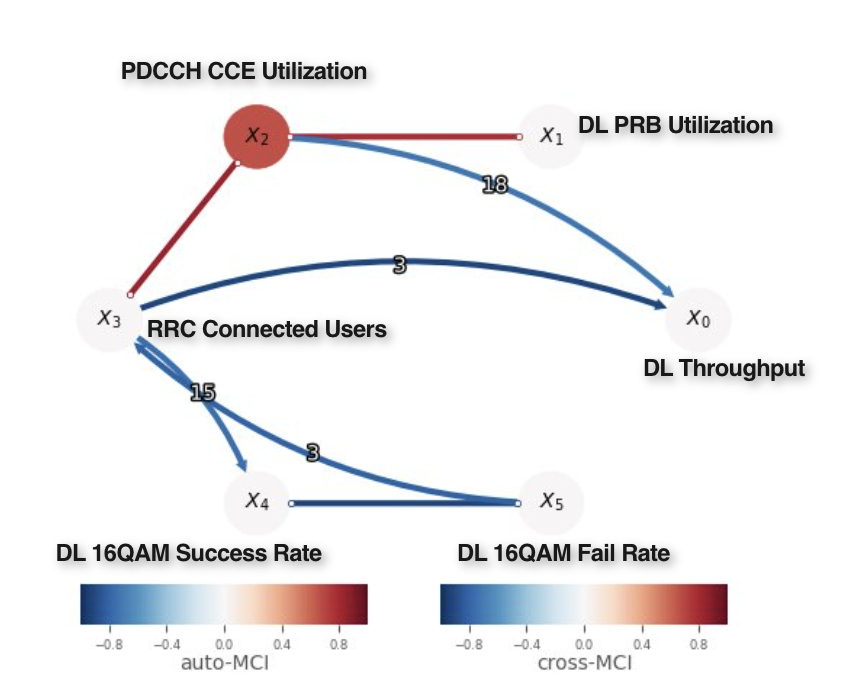}
\caption{\label{fig:PCMCI_normal}PCMCI on Cell Load Issue with Normal State}
\end{figure}

\begin{figure}[htbp!]
\centering
\includegraphics[width=0.85\linewidth]{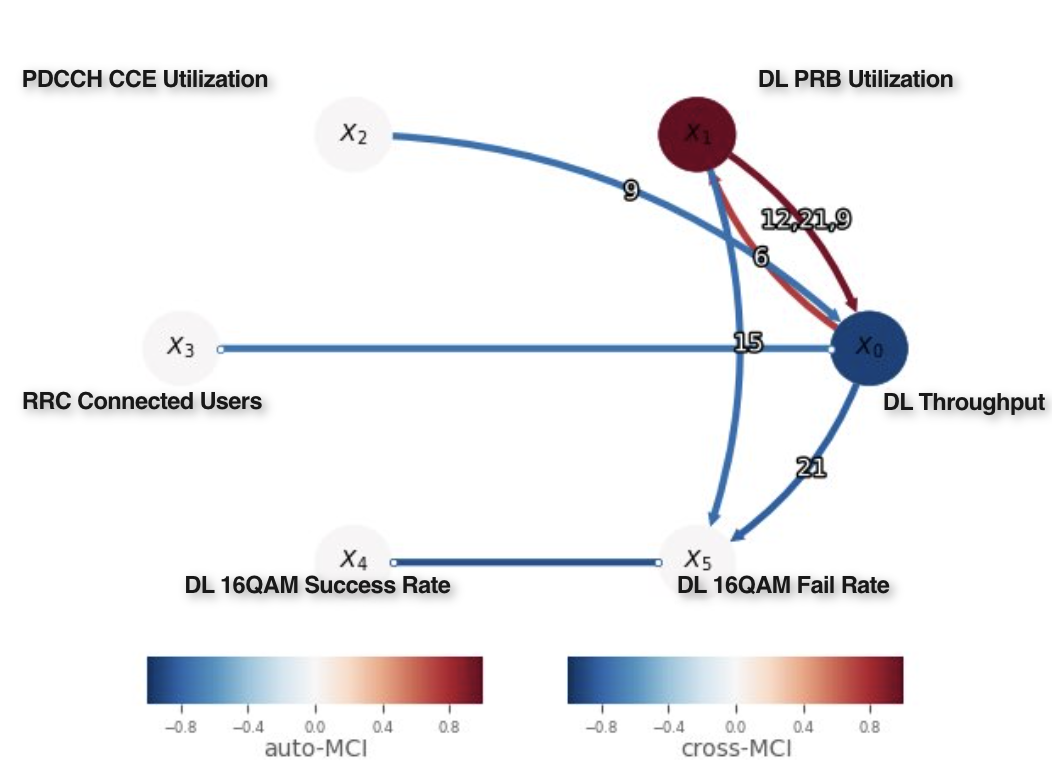}
\caption{\label{fig:PCMCI_abnormal}PCMCI on Cell Load Issue with Abnormal State}
\end{figure}

We start with PCMCI (Peter-Clark Momentary Conditional Independence) as benchmark for causal discovery in telecom use cases, since it is capable of handling  high-dimensional time series data with time-delayed dependencies. However, PCMCI struggles to capture the causal intervention sequence for tracking cell load issues in our experiments, as it does not model counterfactuals or interventional distributions. PCMCI is applied on both normal state and abnormal state so that we could understand the candidates identified as causally relevant root causes and how they deviate from normal behavior in the anomalous state. We could observe that PDCCH CCE Utilization and DL PRB Utilization are leading indicators for load-based DL throughput in Figure~\ref{fig:PCMCI_normal} and Figure~\ref{fig:PCMCI_abnormal}. However, the noise in telecoms data can generate confusion in the outputs, which needs domain knowledge for dimensionality reduction, and new feature synthesis. Additionally, the choice of p-value thresholds, the selection of conditional independence tests (e.g., ParCorr, RobustParCorr), and the configuration of the time lag parameter can lead to varying sets of causal leading indicators, making it challenging for subject matter experts to interpret root causes of network faults.

\vspace{-2pt}

\subsection{RCD + Deviation Detection on Cell Load Issue}
\begin{figure}[htbp!]
\centering
\includegraphics[width=0.8\linewidth]{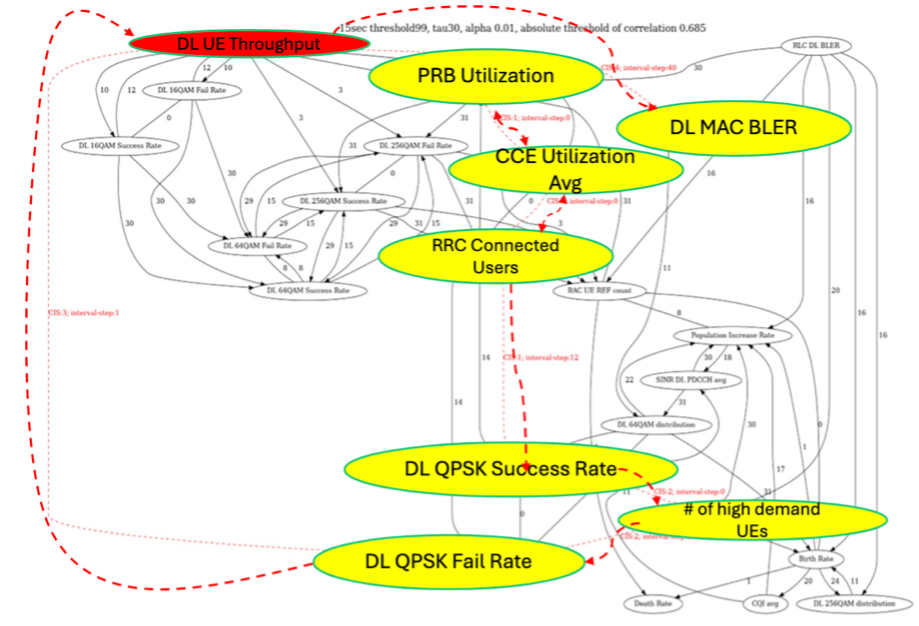}
\caption{\label{fig:CIS}Causally relevant indicators in yellow with SLA breach in red and the preceding intervention events by dashed arrows in red with STEP number in ascending order corresponding to the display sequence of anomaly events.}
\end{figure}

Building on the lessons learned from PCMCI, we propose a combination of AI/ML-driven functional components—namely RCD, the K-S test, and Z-score analysis—to identify the causal intervention sequence. The same dataset used in PCMCI is employed in this approach to ensure a consistent comparison. Results are displayed in Figure~\ref{fig:CIS}. Here, we have configured the number of nodes to be 5 and the number of experiments to be 10 in the RCD algorithm, and the alpha value for CIS p-value to be 0.1 that controls the significance of causal intervention sequence. Nodes and arrows marked in black represent the normal state in the network. Upon SLA breach conditions happening, the preceding events of deviation among causal indicators are highlighted in yellow, and the sequence leading to the throughput degradation below 500kbps is highlighted in red dashed arrows. Here, the RCD algorithm determines the root cause nodes marked in yellow, and the K-S test and the Z-score determine the sequence marked in red dashed arrows. Here, the K-S test identifies the temporal order of causal leading indicators, while the Z-score indicates the direction of deviation from the normal state.

\vspace{-5pt}

\begin{figure}[htbp!]
\centering
\includegraphics[width=0.85\linewidth]{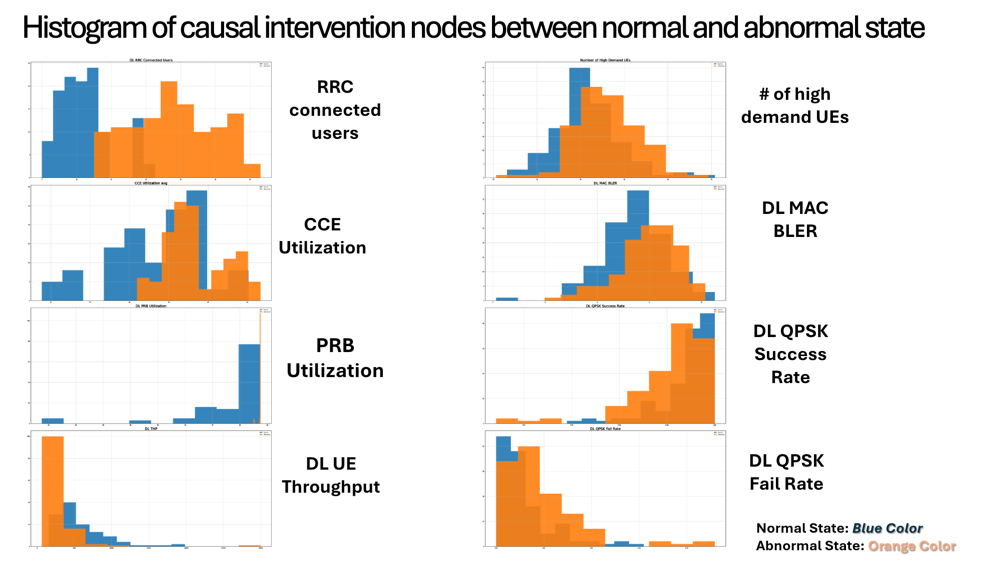}
\caption{\label{fig:histogram}The blue color histogram distribution in normal vs the orange color corresponding to abnormal confirms the results of the method.}
\end{figure}

\begin{figure}[htbp!]
\centering
\includegraphics[width=0.85\linewidth]{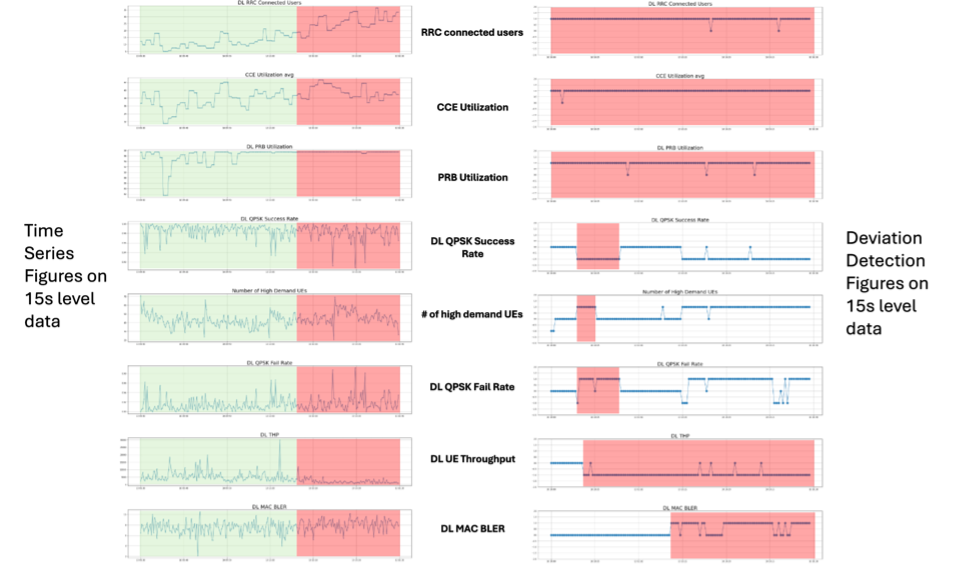}
\caption{\label{fig:deviation}Left panel shows the raw data behavior change from normal in light green to abnormal in purple and the right panel shows the time delay shift of the starting of anomaly events from the output of the univariate deviation detection, where 0 on the right panel corresponds to no anomaly, -1 indicates to decreasing below threshold and 1 indicates increasing above the threshold.}
\end{figure}

To validate the causal intervention sequence, we use the statistical distribution in Figure \ref{fig:histogram}, time series, and deviation detection in Figure \ref{fig:deviation} to validate our results. We can observe that PRB utilization experiences a hard intervention in the abnormal state since it is set to a fixed value, as shown in the 3rd panel from the top on the left half of Figure \ref{fig:histogram}. Remaining KPIs experience soft interventions because their distribution shift in the opposite direction in the abnormal state compared with normal state behavior, implied by the histograms in Figure \ref{fig:histogram}.

\vspace{-5pt}
\vspace{-5pt}

\vspace{-5pt}
\vspace{-5pt}

\subsection{Choice of different p-value under Kolmogorov-Smirnov Test}

\begin{figure}[htbp!]
\centering
\includegraphics[width=0.85\linewidth]{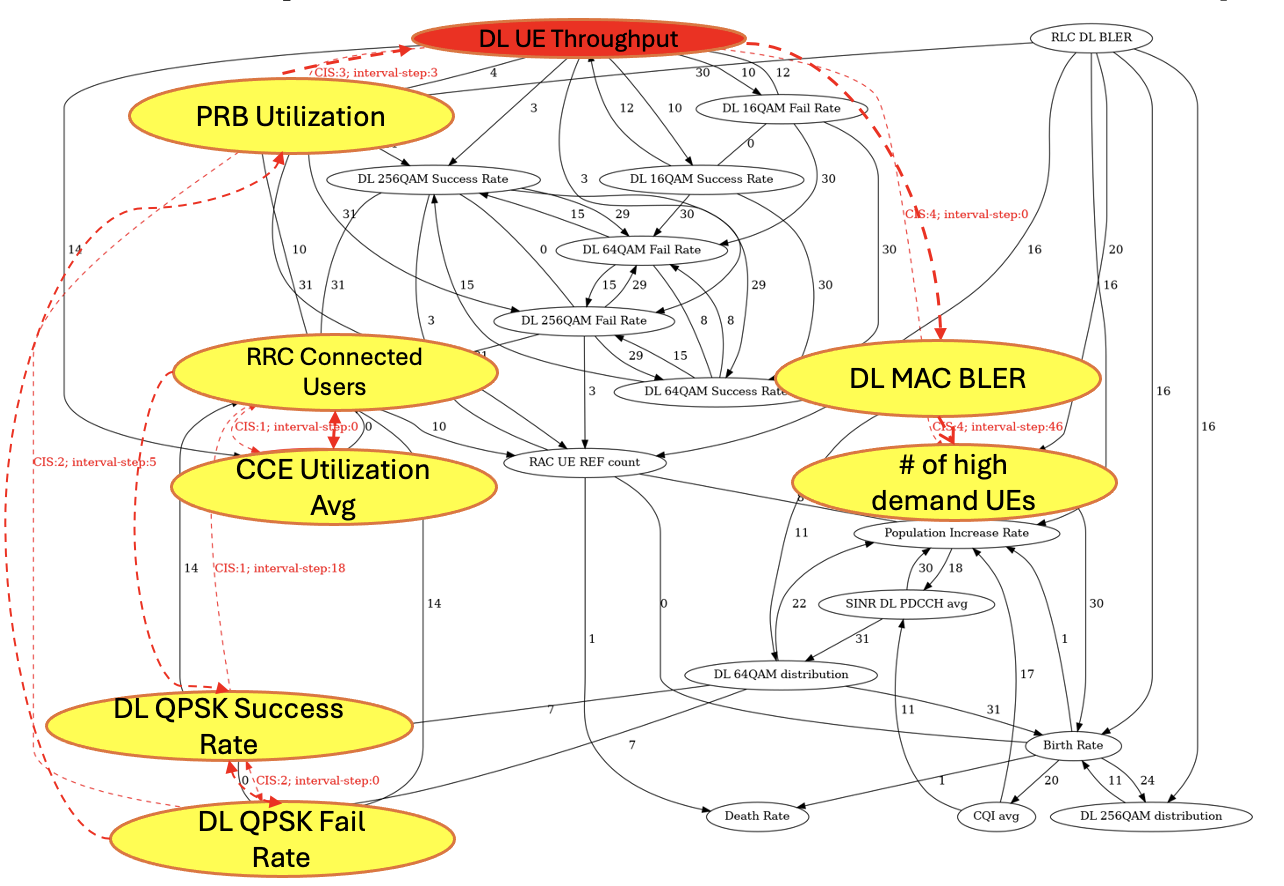}
\caption{\label{fig:pvalue0.05}Casual Intervention Sequence based on the alpha value for CIS p-value to be 0.05.}
\end{figure}

In Figure \ref{fig:CIS}, we have outlined the sequence of causal interventionsons based on the alpha value of the CIS p-value being 0.1. If we reduce CIS p-value to be 0.05, then we will get different casual intervention sequence in Figure \ref{fig:pvalue0.05}. Here, we select a larger CIS p-value to allow greater tolerance for the noisy, high-resolution data observed in the abnormal-state rolling window compared with the normal-state distribution. This choice enables the detection of subtle but consistent shifts in the ordering of causal leading indicators in the telecom domain. To mitigate the increased risk of false positives associated with a higher p-value, we have applied both Bonferroni and FDR corrections, ensuring that statistical significance remains robust.

\vspace{-5pt}

\subsection{Monte Carlo Simulation}
RCD identifies a subset of variables (KPIs) that potentially contribute to the onset of abnormal data distributions. Therefore, the intersection of these subsets, obtained from multiple independent runs of RCD, is expected to reveal the true causal sources. To validate this approach, we employ a Monte Carlo simulation to generate experimental results that complement and support the underlying theoretical analysis of RCD.

\subsubsection{Experimental Setup}
RCD has an input parameter $g$, which needs to randomly pick $g$ nodes for building the causal graphs. The experiment is arranged as follows:
\begin{enumerate}
  \item We consider $g$ as previous, starting from 3 to total number of variables, i.e., 23. 
  \item For each $g$, we have run the simulation experiment in 6 different sets. In the first set, the RCD experiments are run 10 times; in the second to sixth set, the RCD experiments are run 15, 20, 25, 30, 40, and 50 times respectively.
  \item In each set of the run, i.e., for a fixed value of $g$ and an experimental set, we count the number of times each KPI appeared as causal source.
\end{enumerate}

\subsubsection{Experimental Results}
\begin{figure}[htbp!]
\centering
\includegraphics[width=0.66\linewidth]{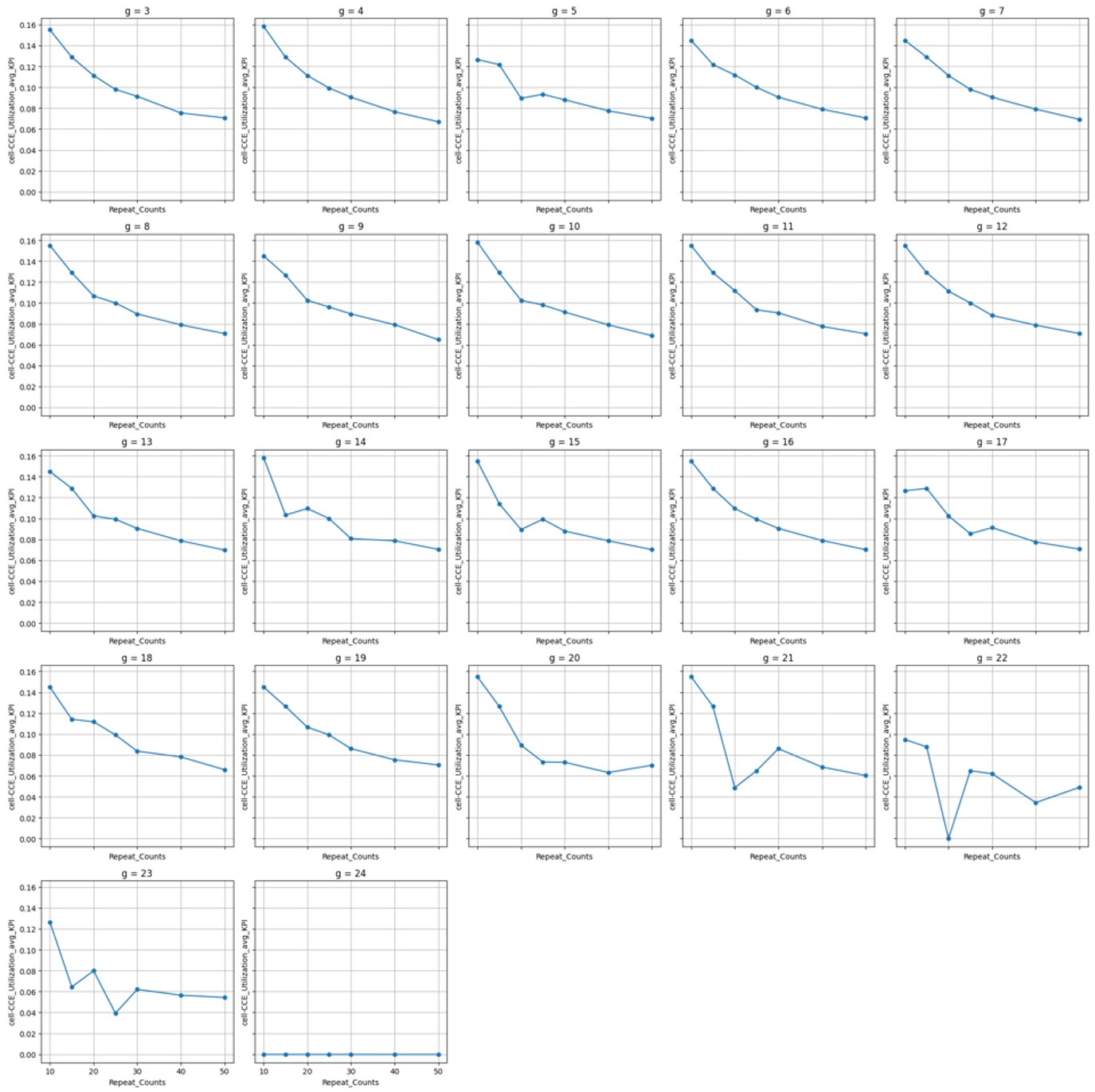}
\caption{\label{fig:mcmc1}Plot of estimated variance of the $P_{g,n}$ vs $n$ for KPI Downlink Throughput}
\end{figure}

For each $(g, n)$ pair, we have collected RCD nodes from the experiment. We are able to find the mean of the value $P_{g,n}$ as the estimate of $p$ for that given $g$. As an example, given below the plot \ref{fig:mcmc1} of estimated variance of the $P_{g,n}$ vs n for the KPI Downlink Throughput. For the prominent causal source KPIs, variability in the estimated $P_{g,n}$ for a given $g$, steadily reduces towards zero as we increase the number of experiments from 10 to 15 to ... 50. Consolidation over all $g$, we first use model based reducing variance trend identification for each $g$ (e.g., Linear Regression). Then, we can choose those KPIs as reliable causal sources if 90\% of the slope of fitted lines are below 0. Hence, we can conclude that we can rely on the RCD findings since the estimate $P_{g,n}$ converges to some value $p$, for all $g$, as $n$ increases.

Next, we aim to quantify and estimate the likelihood that a given KPI is a causal source. For a fixed value of $g$, we compute the mean of the corresponding values 
$P_{g,n}$, which serves as an estimate of the $p$. To determine the optimal value of $g$, we evaluate the estimated $p$ across a range of $g$ values from 3 to 23. The median of these estimates is then selected, and the associated $g$ is proposed as the optimal value for RCD. In the case of an even number of estimates resulting in two median values, we select the smaller $g$ as the proposed value. To identify the optimal value of $n$, we analyze the proportion of variance reduction for the selected $g$, varying $n$ over the set {10, 15, ..., 30, 40, 50}. The value of $n$ that yields the greatest reduction in variance is chosen. The final values of $g$, $n$, and the estimated probability of each KPI being a causal source are summarized in the following table \ref{tab:kpi_probability_optimaln}.

\begin{table}[htbp!]
\centering
\caption{Parameter $g$, Estimated Probability, and Optimal $n$ for KPIs}
\begin{tabular}{|>{\centering\arraybackslash}p{3.2cm}|
                >{\centering\arraybackslash}p{1.2cm}|
                >{\centering\arraybackslash}p{1.5cm}|
                >{\centering\arraybackslash}p{1.2cm}|}
\hline
\textbf{KPI Name} & \textbf{Parameter $g$} & \textbf{Probability Estimation} & \textbf{Optimal $n$} \\
\hline
DL\_QPSK\_Fail\_Rate & 3 & 0.14 & 20 \\
\hline
DL\_16QAM\_Success\_Rate & 3 & 0.02 & 40 \\
\hline
DL\_QPSK\_Success\_Rate & 4 & 0.16 & 0 \\
\hline
DL\_QPSK\_Distribution & 3 & 0.00 & 30 \\
\hline
DL\_256QAM\_Distribution & 3 & 0.02 & 0 \\
\hline
DL\_16QAM\_Fail\_Rate & 3 & 0.02 & 50 \\
\hline
RAC\_UE\_REF\_Death\_Rate & 3 & 0.00 & 0 \\
\hline
RRC\_Connected\_Users\_DL & 3 & 1.00 & 0 \\
\hline
SINR\_DL\_PDCCH\_AVG & 3 & 0.00 & 0 \\
\hline
DL\_64QAM\_Success\_Rate & 3 & 0.14 & 40 \\
\hline
DL\_64QAM\_Distribution & 3 & 0.00 & 0 \\
\hline
DL\_16QAM\_Distribution & 3 & 0.02 & 0 \\
\hline
DL\_256QAM\_Fail\_Rate & 5 & 0.14 & 0 \\
\hline
DL\_64QAM\_Fail\_Rate & 3 & 0.15 & 15 \\
\hline
MAC\_DL\_BLER & 4 & 0.50 & 0 \\
\hline
DL\_256QAM\_Success\_Rate & 4 & 0.12 & 0 \\
\hline
CCE\_Utilization\_AVG & 3 & 0.45 & 20 \\
\hline
RLC\_DL\_BLER & 3 & 0.00 & 0 \\
\hline
\end{tabular}
\label{tab:kpi_probability_optimaln}
\end{table}

Table \ref{tab:kpi_probability_optimaln} summarizes the result of a Monte Carlo simulation experiment. It is apparent that  a KPI will be a poor causal source if n = 0 and p \textless\:1. The selection of  prominent causal sources can be designed as per user requirement. A set of the prominent causal sources may be chosen by thresholding on estimated $p$ and $n$, e.g., n \textgreater\:0, estimated p \textgreater\:0.4, or choose 5\% causal sources by rank ordering $p$ where n \textgreater\:0. Thus, RRC\_Connected\_Users\_DL and CCE\_Utilization\_AVG are prominent causal sources. Hence, We have an approach to identify the adequate Monte Carlo experimental setup for RCD. Meanwhile, we have estimated the probabilities that individual KPIs are causal sources and enhance rank ordering of prominent causal sources.

\section{Conclusion}
We introduced an AI/ML framework that pinpoints the root causes of Service-Level Agreement (SLA) violations in Radio Access Networks without the heavy manual effort that usually slows operators down. The workflow blends three lightweight modules—Root-Cause Discovery (RCD), causal sub-graph analysis, and deviation detection—to trace a clear chain of events from the first anomaly to the SLA breach. Because the entire pipeline runs on standard CPUs and samples only a fraction of the input variables, it remains energy- and cost-efficient even when monitoring thousands of cells. Monte-Carlo simulations confirm that accuracy holds as the network scales, while a simple histogram-based check gives engineers an intuitive, visual way to validate each flagged indicator. In short, the framework turns high-volume telemetry into actionable insights that engineers can trust—helping them resolve faults faster and keep subscribers connected. 


\end{document}